\documentclass[aps,showpacs,preprint]{revtex4}

\usepackage{amssymb}

\usepackage{graphicx}

\begin{document}

\title{Quantum theory of SASE FEL}
\author{R. Bonifacio$^{a}$, N.\ Piovella$^{a}$, G.R.M.\ Robb$^{b}$}
\affiliation{ $^{a}$Dipartimento di Fisica, Universit\`a Degli
Studi di Milano, INFM and INFN, Via Celoria 16, I-20133 Milano, Italy.\\
$^{b}$Department of Physics, University of Strathclyde, Glasgow,
G4 0NG, Scotland.}

\begin{abstract}
We describe a free-electron laser (FEL) in the Self Amplified
Spontaneous Emission (SASE) regime quantizing the electron motion
and taking into account propagation effects. We demonstrate
quantum purification of the SASE spectrum, i.e., in a properly
defined quantum regime the spiking behavior disappears and the
SASE power spectrum becomes very narrow.
\end{abstract}
 \maketitle
%%%%%%%%%%%%%%%%%%%%%%%%%%%%%%%%%%%%%%%%%%%%%%%%%%%%%%%%%%%%%

%%%%%%%%%%%%%%%%%%%%%%%%%%%%%%%%%%%%%%%%%%%%%%%%%%%%%%%%%%%%%

The Self Amplified Spontaneous emission (SASE) regime for a
free-electron laser (FEL) is made up of three basic ingredients:
high-gain, propagation or ``slippage'' effects and start-up from
noise \cite{SASE}. The classical steady-state high-gain regime of
FELs, with universal scaling and the introduction of the
$\rho$-BPN parameter, was analysed in ref.\cite{BPN}, where the
possibility of operating an FEL in the SASE regime was described.
The first experimental observation of the high-gain regime, also
starting from noise, was carried out in the microwave range using
a waveguide in the Livermore experiment \cite{Livermore}.
Presently, short wavelength FELs which amplify incoherent shot
noise via SASE are of great interest worldwide as potential
sources of ultra bright coherent X-ray radiation
\cite{SASE:exp1,SASE:exp2}.

Many theoretical studies of high-gain FELs \cite{altri} do not
take into account propagation effects and the initial noise is
described by a small input signal or a small bunching. Other
treatments assume that SASE is just steady-state amplification
starting from noise, ignoring propagation effects \cite{KJK,Wang}.
That approach does not give the correct temporal structure and
spectrum of the SASE radiation as described in ref. \cite{SASE}.

In ref. \cite{FEL:SR,NC} it was shown that due to propagation
there exists not only the steady-state instability of ref. \cite{BPN},
but also a superradiant instability, with peak intensity
proportional to $n^2$, where $n$ is the electron density. This
instability originates in the region near the rear edge of the
electron bunch, producing a soliton-like pulse which grows and narrows as it
slips over the electrons, preserving a hyperbolic secant shape
with some ringing \cite{self similar}. We stress that the
mathematics and the physics of this superradiant instability,
which is at the heart of SASE \cite{SASE}, is completely different
from the usual steady-state instability. A striking example of the difference
between the steady-state and superradiant regimes is that
superradiantly amplified spikes occur when the system is detuned
from resonance, whereas the steady-state instability does not occur
\cite{FEL:SR,NC}. In this case, the superradiant instability is
actually stronger than for the case of exact resonance, since the
radiation pulse propagates over fresh electrons, which have not
been perturbed by the steady-state instability.

As shown in ref. \cite{SASE}, a SASE FEL radiates a random series
of spikes since, roughly speaking, the electron bunch contains
many cooperation lengths $L_c$ which radiate randomly and
independently from one another. The final result is an almost
chaotic temporal pulse structure with a broad spectral width. The
number of spikes in the high-gain regime corresponds approximately
to the number of cooperation lengths in the electron bunch. Only
when the length of the electron bunch is a few cooperation
lengths is a partial ``cleaning'' of the temporal profile of the
radiation obtained, as described in ref. \cite{SASE}.
Furthermore, the total radiated energy does not behave as that
predicted by steady-state theory. After reaching the
steady-state value of saturation, the energy for SASE continues to
increase due to the fact that the superradiant spikes do not
saturate. However, SASE has one drawback with regard to its application
as a useful source of short-wavelength coherent light : at short
wavelengths many cooperation lengths $L_c$ lie within the electron
bunch. This implies a quasi-chaotic temporal structure of the
radiation pulse and a consequent wide spectrum.

Several schemes have been proposed in order to avoid the large
spectral width associated with classical SASE. One of these
involves a multiple wiggler scheme with a coherent seed laser
\cite{harm1,harm2}. This method is based on the fact that the FEL
interaction creates a strong bunching not only on the fundamental
but also on the higher harmonics (even and odd), as shown
analytically in \cite{harm1,Corsini}. Hence, if the electron beam
is injected in a second wiggler tuned to one of the higher
harmonics, even or odd, the system starts radiating with intensity
$\propto n^2$. This superradiant harmonic generation (SRHG)
eventually evolves into the high-gain exponential regime. A
difficulty in producing SRHG is that ``it is necessary to make the
first wiggler section long enough to bring the harmonic bunching
well above noise, but not so long as to induce too much energy
spread'' \cite{harm2}. This optimization of SRHG, termed high-gain
harmonic generation (HGHG), has recently been performed by other
authors \cite{LHua}. However, this method presents difficulties
when going to very short wavelengths, as pointed out in ref.
\cite{Saldin}, due to the fact that each stage amplifies not only
the coherent signal but also the noise.

In this letter we propose a novel method for producing coherent
short wavelength radiation with SASE. We introduce a quantum
description of SASE which depends on a dimensionless quantum FEL
parameter, $\bar\rho$, which determines the number of photons per
electron and/or the electron recoil in units of the photon
momentum. We show that when $\bar\rho \gg 1$ the SASE FEL behaves
classically, i.e. in agreement with the SASE classical model.
However, when $\bar\rho\le 1$ we obtain a quantum regime with
features completely different from those of the classical regime,
and which we shall refer as Quantum SASE. A surprising feature of
this regime is the phenomenon of ``quantum purification'', in
which the random spiking behavior almost disappears and a strong
narrowing of the spectrum occurs. Here we generalize a previous
model \cite{new} including propagation effects via a multiple
scaling method used in the classical FEL theory of ref.
\cite{Scully}. This allows us to easily take into account the
existence of two different spatial scales in the FEL: the
variation of the electron distribution on the scale of the
radiation wavelength (describing the bunching) and the variation
of the field envelope on the much longer scale of the cooperation
length.

The quantum FEL is described by the following equations for the
dimensionless radiation amplitude $A(\bar z,z_1)$ and the matter
wave field $\Psi(\theta,\bar z,z_1)$:
\begin{eqnarray}
i\frac{\partial \Psi (\theta ,\bar{z},z_{1})}{\partial \bar{z}}
&=&-\frac{1}{ 2\bar{\rho}}\frac{\partial ^{2}}{\partial \theta
^{2}}\Psi (\theta, \bar{z},z_{1})-i\bar{\rho}\left[
A(\bar{z},z_{1})e^{i\theta }-\mathrm{
c.c.}\right] \Psi (\theta ,\bar{z},z_{1}) \label{MS1}\\
\frac{\partial A(\bar{z},z_{1})}{\partial \bar{z}}+\frac{\partial
A(\bar{z},z_{1})}{\partial z_{1}} &=&\frac{1}{2\pi }\int_{0}^{2\pi
}\;d\theta |\Psi (\theta ,\bar{z},z_{1})|^{2}e^{-i\theta
}+i\bar\delta A(\bar{z},z_{1})\label{MS2}.
\end{eqnarray}
These equations were derived in detail in ref.\cite{new}, to
describe Collective Recoil Lasing (CRL) in FELs and in atomic
systems. In the CRL equations,  the electrons are described by a
Schr\"{o}dinger equation for a matter-wave field $\Psi$ \cite
{PREP,EPL} in a self-consistent pendulum potential proportional to
$A$, where $|A|^2=|a|^2/(N\bar\rho)$, $|a|^2$ is the average
number of photons in the interaction volume $V$, and $|\Psi|^2$ is
the space-time dependent electron density, normalized to unity. In
Eqs. (\ref{MS1}) and (\ref{MS2}) we have adopted the universal
scaling used in the classical FEL theory \cite{BPN,SASE,NC},
\textit{i.e.} $\theta=(k+k_w)z-ckt$ is the electron phase, where
$k_w=2\pi/\lambda_w$ and $k=\omega/c=2\pi/\lambda$ are the wiggler
and radiation wavenumbers, $\bar z=z/L_g$ is the dimensionless
wiggler length, $L_g=\lambda_w/4\pi\rho$ is the gain length,
$z_1=(z-v_rt)/\beta_rL_c$ is the coordinate along the electron
bunch moving at the resonant velocity $v_r=c\beta_r=ck/(k+k_w)$
and $L_c=\lambda/4\pi\rho$ is the cooperation length or coherence
length, $\rho=\gamma_r^{-1}(a_w/4ck_w)^{2/3}(
e^2n/m\epsilon_0)^{1/3}$ is the classical BPN parameter
\cite{BPN}, $\gamma_r=\sqrt{(\lambda/2\lambda_w)(1+a_w^2)}$ is the
resonant energy in $mc^2$ units, $a_w$ is the wiggler parameter
and $n$ is the electron density. Finally, $\bar
p=(\gamma-\gamma_0)/\rho\gamma_0$ is the dimensionless electron
momentum and $\bar\delta=(\gamma_0-\gamma_r)/\rho\gamma_0$ is the
detuning parameter, where $\gamma_0\approx\gamma_r$ is the initial
electron energy in $mc^2$ units. It can be easily shown that Eq.
(\ref{MS1}) implies:
\begin{equation}\label{norm}
\frac{\partial }{\partial \bar{z}}\int_{0}^{2\pi }d\theta \left|
\Psi (\theta ,z_{1},\bar{z})\right| ^{2}=0.
\end{equation}
Hence, the dimensionless density profile
$I_{0}(z_{1})=\int_{0}^{2\pi }d\theta |\Psi |^{2}$ is independent
of $\bar{z}$. This means that the spatial distribution of the
particles does not change appreciably on the slow scale $z_{1}$
during the interaction with the radiation.

Whereas the classical FEL equations in the above universal scaling
do not contain any explicit parameter (see ref. \cite{FEL:SR,NC}),
the quantum FEL equations (\ref{MS1}) and (\ref{MS2}) depend on
the quantum FEL parameter
\begin{equation}\label{rhobar}
    \bar\rho=\left(\frac{mc\gamma_r}{\hbar k}\right)\rho.
\end{equation}
From the definition of $A$, it follows that
$\bar\rho|A|^2=|a|^2/N$ is the average number of photons emitted
per electron. Hence, since in the classical steady-state high-gain
FEL $A$ reaches a maximum value of the order of unity, $\bar\rho$
represents the maximum number of photons emitted per electron, and
the classical regime occurs for $\bar\rho\gg 1$. Note also that in
Eq. (\ref{MS1}) $\bar\rho$ appears as a ``mass'' term, so one
expects a classical limit when the mass is large. As we shall see,
when $\bar\rho < 1$ the dynamical behavior of the system changes
substantially from a classical to a quantum regime.

A further inspection of Eq. (\ref{MS1}) shows that $\Psi$ depends
explicitly on $\theta$, which describes the variation (bunching)
on the scale of the radiation wavelength $\lambda$. The
parametrical dependence on $z_1$ is induced by the slow
spatio-temporal evolution of the field amplitude $A$. This
evolution is described by Eq. (\ref{MS2}), in which the bunching
factor is the ensemble average of $e^{-i\theta}$. In other words,
the $N$-particle system is described as a quantum ensemble
represented by the matter-wave field $\Psi$. This model has been
used previously to describe the quantum regime of an FEL
\cite{PREP} and of the Collective Atomic Recoil Laser (CARL)
\cite{CARL1,CARL2,CARL3,Moore:1,Gatelli,Cola}, neglecting the
dependence on $z_1$, i.e., propagation. This is appropriate for
the FEL when slippage due to the difference between the light and
electron velocities is neglected, which is never true in the SASE
regime where the propagation from one cooperation length to
another is fundamental.

The classical limit of the FEL can be explicitly shown as follows.
Eq.(\ref{MS1}) can be transformed into an equation for the Wigner
function, as shown in ref. \cite{EPL}:
\begin{equation}
\frac{\partial W(\theta ,\bar{p},z_{1},\bar{z})}{\partial \bar{z}}
+\bar{p} \frac{\partial W(\theta ,\bar{p},z_{1},\bar{z})}{\partial
\theta }-\bar{\rho}\left( Ae^{i\theta }+A^{\ast }e^{-i\theta
}\right) \left[ W\left( \theta
,\bar{p}+\frac{1}{2\bar{\rho}},z_{1},\bar{z}\right) -W\left(
\theta ,\bar{p}-\frac{1}{2\bar{\rho}},z_{1},\bar{z}\right) \right]
=0,\label{MW1}
\end{equation}
whereas Eq.(\ref{MS2}) becomes
\begin{equation}
\frac{\partial A}{\partial \bar{z}}+\frac{\partial A}{\partial
z_{1}} = \frac{1}{2\pi }\int_{-\infty }^{+\infty
}d\bar{p}\int_{0}^{2\pi }d\theta \,W(\theta
,\bar{p},z_{1},\bar{z})e^{-i\theta }+i\bar\delta A. \label{MW2}
\end{equation}
In the right hand side of Eq. (\ref{MW1}), the incremental ratio
$\bar\rho[W(\theta ,\bar{p}+1/2\bar{\rho},z_{1},\bar{z} )-W(\theta
,\bar{p}-1/2\bar{\rho},z_{1},\bar{z})]\rightarrow
\partial W(\theta ,\bar{p},z_{1},\bar{z})/\partial \bar{p}$ when $\bar{\rho}
\rightarrow \infty $. Hence, for large values of $\bar{\rho}$, Eq.
(\ref{MW1}) becomes the Vlasov equation:
\begin{equation}
\frac{\partial W(\theta ,\bar{p},z_{1},\bar{z})}{\partial
\bar{z}}+\bar{p} \frac{\partial W(\theta
,\bar{p},z_{1},\bar{z})}{\partial \theta }-\left( Ae^{i\theta
}+A^{\ast }e^{-i\theta }\right) \frac{\partial W(\theta ,\bar{p}
,z_{1},\bar{z})}{\partial \bar{p}}=0.  \label{Vlasov}
\end{equation}
Eqs. (\ref{MW1}) and (\ref{MW2}) provide the description of the
CRL model in terms of the Wigner function for the electrons,
whereas  Eqs. (\ref{MW2}) and (\ref{Vlasov}) are equivalent to the
classical FEL model of ref. \cite{Scully}. Note that Eqs.
(\ref{MW2}) and (\ref{Vlasov}) do not depend explicitly on
$\bar\rho$, as must be the case in the classical model with
universal scaling \cite{FEL:SR,NC}. We briefly mention that
Eq.~(\ref{MW1}) for the Wigner function has a broader validity
than the Schr\"{o}dinger equation (\ref{MS1}), since it can also
describe a statistical mixture of states which cannot be
represented by a wave function but rather by a density operator.

\begin{figure}
\begin{center}
\includegraphics[width=14truecm]{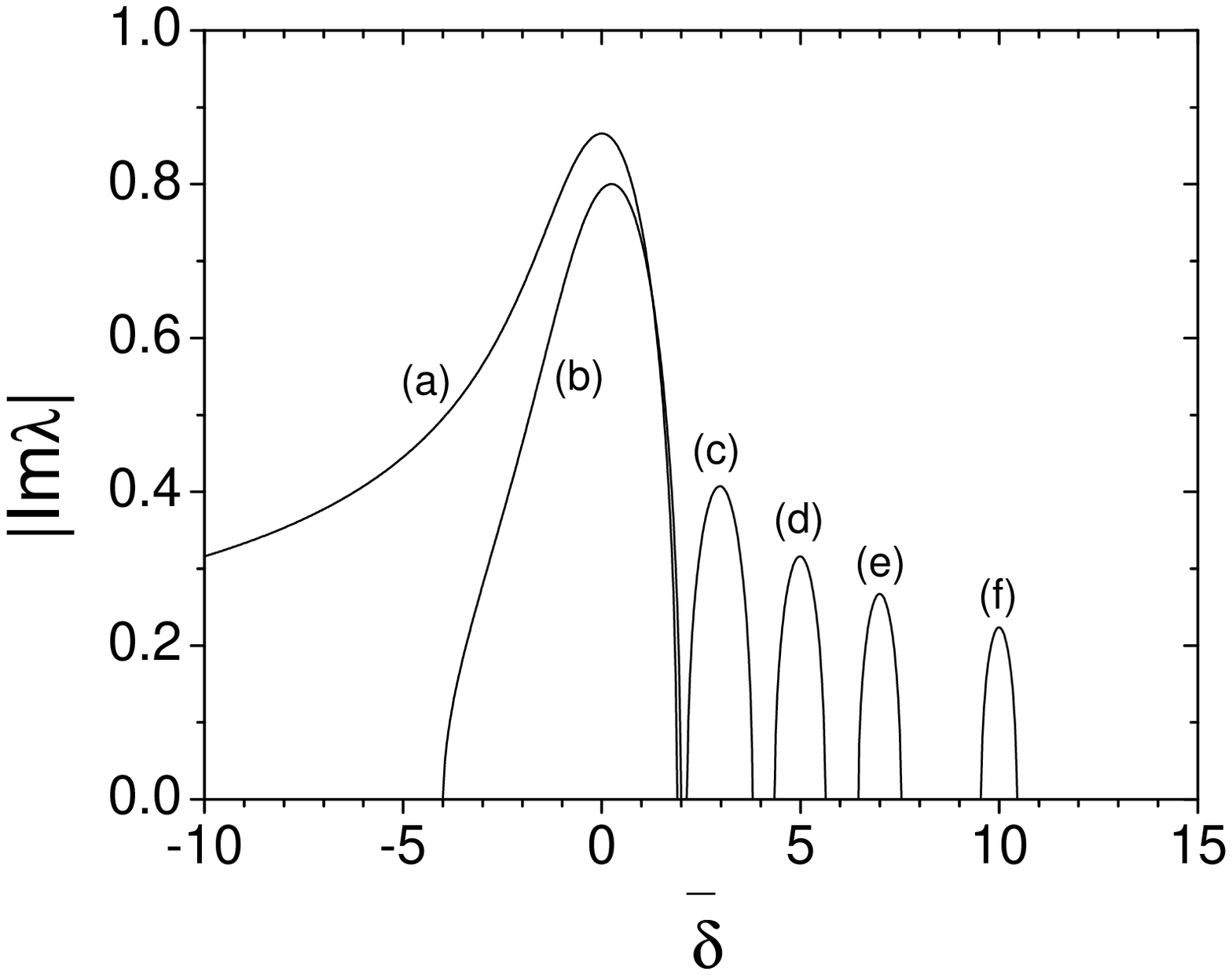}
\end{center}
\caption{Imaginary part of the unstable root of the cubic equation
(\ref {cubica}) vs. $\bar\delta$, for $1/2\bar{\protect\rho} =0$,
(a), $0.5$, (b), $3$, (c), $5$, (d), $7$, (e) and $10$, (f).}
\label{fig1}
\end{figure}

Eqs.(\ref{MS1}) and (\ref{MS2}) are conveniently solved in the
momentum representation. Assuming that $\Psi (\theta
,z_{1},\bar{z})$ is a periodic function of $\theta $, it can be
expanded in a Fourier series:
\begin{equation}
\Psi (\theta ,z_{1},\bar{z})=\sum_{n=-\infty }^{\infty
}c_{n}(z_{1},\bar{z})e^{in\theta }  \label{Fourier}
\end{equation}
so inserting Eq. (\ref{Fourier}) into Eqs. (\ref{MS1}) and (\ref{MS2}), we
obtain
\begin{eqnarray}
\frac{\partial c_{n}}{\partial \bar{z}}
&=&-\frac{in^{2}}{2\bar{\rho}}c_{n}-
\bar{\rho}\left( Ac_{n-1}-A^{\ast }c_{n+1}\right)  \label{cn1} \\
\frac{\partial A}{\partial \bar{z}}+\frac{\partial A}{\partial
z_{1}} &=&\sum_{n=-\infty }^{\infty }c_{n}c_{n-1}^{\ast
}+i\bar\delta A,  \label{cn2}
\end{eqnarray}
Eqs. (\ref{cn1}) and (\ref{cn2}) are our working equations and
their numerical analysis will be discussed in the following. Note
that from Eq. (\ref{Fourier}) it follows that $|c_{n}|^{2}$ is the
probability that an electron has a dimensionless momentum $\bar
p=n/\bar\rho$ (i.e. $mc(\gamma-\gamma_0)=(\hbar k)n$),
\begin{equation}
b=\sum_{n=-\infty }^{\infty }c_{n}c_{n-1}^{\ast }  \label{bunching}
\end{equation}
is the quantum expression of the bunching parameter and
\begin{equation}
\langle\bar p\rangle=\frac{1}{\bar\rho}\sum_{n=-\infty }^{\infty
}n|c_{n}|^{2} \label{ave p}
\end{equation}
is the average dimensionless momentum. Note that the quantum
bunching (\ref{bunching}) requires a coherent superposition of
different momentum states. The stability analysis of Eqs.
(\ref{cn1}) and (\ref {cn2}) has been carried out in ref.
\cite{Moore:1,Gatelli,Cola} for the case with no propagation. We
assume that the system is in an equilibrium state with no field,
$A=0$, and all the electrons are in the state $n=0$, with $c_{0}=1$
and $c_{m}=0$ for all $m\neq 0$. This equilibrium state is
unstable when the dispersion relation
\begin{equation}
(\lambda -\bar\delta) \left( \lambda ^{2}-\frac{1%
}{\bar{4\rho}^{2}}\right) +1=0,  \label{cubica}
\end{equation}
has complex roots. Notice that this dispersion relation coincides
with that of a classical FEL with an initial energy spread with a
square distribution and width $1/\bar\rho$ \cite{NC}, i.e., this
extra term represents the intrinsic quantum momentum spread which,
in dimensional units, becomes $\hbar k$. In fig.\ref{fig1} we plot
the imaginary part of $\lambda$ as a function of $\bar\delta$ for
different values of $\bar{\rho}$. The classical limit is obtained
for $\bar{\rho}\gg 1$ (see fig.\ref{fig1}a). In this case, the
system is unstable for $\bar\delta \lesssim 2$, with maximum
instability rate $\text{Im}\lambda =\sqrt{3}/2$ at $\bar\delta
=0$. When $\bar{\rho}$ is smaller than unity (fig.\ref{fig1}c-f),
the instability rate decreases and the peak of
$\mathrm{Im}(\lambda )$ moves around $\bar\delta =1/2\bar\rho$
(\textit{i.e.} $mc(\gamma _{0}-\gamma _{r})=\hbar k/2$) with peak
value $\text{Im}\lambda =\sqrt{\bar{\rho}}$ and full width on
$\bar\delta$ equal to $4\bar{\rho}^{1/2}$ (i.e.
$mc\Delta\gamma=4(\hbar k)\bar\rho^{3/2}$). As discussed in ref.
\cite {Gatelli,Cola}, for $\bar{\rho}\gg 1$ the electrons have
almost the same probability of transition from the momentum state
$n=0$ to $n=1$ or $n=-1$ (i.e. $|c_{1}|^{2}\sim |c_{-1}|^{2}$),
absorbing or emitting a photon. On the contrary, in the case
$\bar{\rho}\leq 1$, $|c_{1}|^{2}\ll |c_{-1}|^{2}$, i.e. the
particles can only emit a photon without absorption.

As shown by the linear stability analysis discussed above, the
quantum regime occurs for small value of $\bar{\rho}$, when an
electron emits only a single photon. In this limit, the dynamics
of the interaction is that of a system with only two momentum
states, i.e. the initially occupied state with $n=0$ and the
recoiling state with $n=-1$. In this limit, Eq. (\ref{cn1} ) and
(\ref{cn2}), after defining the ``polarization''
$S=c_{0}c_{-1}^{\ast }\exp [i\bar\delta\bar{z}]$ and the
``population difference'' $D=|c_{0}|^{2}-|c_{-1}|^{2}$, reduce to
the so-called ``Maxwell-Bloch equations'' for a two-state system
\cite{MBE}:
\begin{eqnarray}
\frac{\partial }{\partial \bar{z}}S(z_{1},\bar{z}) &=&-i\Delta
S(z_{1},\bar{z})+\bar{\rho}\bar{A}(z_{1},\bar{z})D(z_{1},\bar{z})  \label{MB1} \\
\frac{\partial }{\partial \bar{z}}D(z_{1},\bar{z})
&=&-2\bar{\rho}\left[ \bar{A}(z_{1},\bar{z})^{\ast }S(z_{1},\bar{z})+\text{c.c.}\right]  \label{MB2} \\
\frac{\partial \bar{A}}{\partial \bar{z}}+\frac{\partial
\bar{A}}{\partial z_{1}} &=&S(z_{1},\bar{z}).  \label{MB3}
\end{eqnarray}
where $\Delta =\bar\delta -1/2\bar{\rho}$ and $\bar{A}=A\exp
{(-i\bar\delta\bar{z})}$. Notice that the parameter $\bar{\rho}$
in Eqs. (\ref{MB1})-(\ref{MB3}) may
be eliminated by redefining the variables as $A^{\prime }=\sqrt{\bar{\rho}}\bar{%
A}$, $z^{\prime }=\sqrt{\bar{\rho}}\bar{z}$, $z_{1}^{\prime }=\sqrt{\bar{\rho%
}}z_{1}$ and $\Delta ^{\prime }=\Delta /\sqrt{\bar{\rho}}$. With
this quantum universal scaling the cooperation length becomes
$L_{c}^{\prime }=L_{c}/\sqrt{\bar{\rho}}\propto 1/\sqrt{n}$.
An interesting result of this scaling is that the CRL model of Eqs. (\ref{MS1})
and (\ref{MS2}) can now be interpreted as a Schr\"{o}dinger equation
for a single particle with a ``mass'' $\bar{\rho}^{3/2}$ in a
self-consistent pendulum potential. This provides an intuitive
interpretation of the classical limit which holds when the
particle's ``mass'' is large.

We now discuss some analytical results arising from the quantum SASE model.
It has been shown in a previous work \cite{new} that the classical
model of Eqs. (\ref{MW2}) and (\ref{Vlasov}) admits a soliton-like
self-similar solution $A(\bar z,z_1)=z_1A_1(y)$ when
$\bar\delta=0$, where $y=\sqrt{z_1}(\bar
z-z_1)\approx(z-v_rt)^{1/2}(ct-z)/L_c^{3/2}$ (where we have
assumed $\beta_r\sim 1$) \cite{self similar}. This means that the
radiation pulse propagates over the electron bunch (i.e. at
different values of $z_1$), preserving its shape but increasing
with amplitude $\propto z_1$ and narrowing with width $\propto
1/\sqrt{z_1}$. It is possible to show \cite{NC} that
in the linear regime the radiation pulse has a maximum at
$z_1=\bar z/3$ and propagates at a constant velocity
$v_s=3v_r/(2+\beta_r)$.

An analogous self-similar solution also exists for the quantum
equations (\ref{MB1})-(\ref{MB3}) when $\Delta=0$ \cite{Burn},
i.e. $A(\bar z,z_{1})=z_1A_2(x)$, where $x=\bar\rho z_1(\bar
z-z_1)=(z-v_rt)(ct-z)/L_c'^2$ (with $L_c'=L_c/\sqrt{\bar\rho}$).
The shape of the radiation pulse is similar in the classical and
quantum cases, but in the quantum case its width decreases as
$1/\bar\rho z_1$. In this case, the radiation pulse in the linear
regime has a maximum at $z_1=\bar z/2$ and moves at the constant
velocity $v_s=2v_r/(1+\beta_r)$.

From the features of the self-similar solutions we can deduce some
important aspects of the nature of the random spikes emitted in
the SASE regime. First, both in the classical and in the quantum
regime the dimensionless amplitude $A$ of the field is
proportional to $z_1\propto \rho\propto N^{1/3}$, so that the
number of emitted photons $|a|^{2}=N\bar{\rho}|A|^{2}\propto
N^{2}$, ie. is superradiant. Whereas the characteristic spatial
length in the classical regime is $L_c$, in the quantum regime the
characteristic length is $L_c'=L_c/\sqrt{\bar\rho}\gg L_c $ for
$\bar\rho\ll 1$.

Let us now briefly restate the reasons for the classical random
spiking behavior. If the bunch length contains many cooperation
lengths $L_c$, each of them radiates a superradiant spike
independently as the electron bunch propagates into the wiggler.
Each spike is only roughly represented by a self-similar solution
because the radiation from one portion does not find fresh
electrons but electrons which have already interacted with radiation. As
discussed in \cite{SASE} the number of spikes increases with the
number of cooperation length in the bunch, i.e. $L_b/2\pi L_c$ .

We now describe the numerical solution of Eqs.(\ref{MS1}) and
(\ref{MS2}) which demonstrates the different dynamical behavior in
the classical and quantum SASE regimes. The SASE simulation has
been performed assuming all the electrons are initially in the
$n=0$ momentum state and that there is a weak randomly fluctuating
modulation in the electron density along the electron pulse, as is
appropriate to model random electron shot noise. The initial
conditions for all the simulations are therefore
$A(z_1,\bar{z}=0)=0$, $c_{-1}(z_1,\bar{z}=0)=b_0 e^{i \phi(z_1)}$
and $c_0(z_1,\bar{z}=0)=\sqrt{1 - b_0^2}$, where $b_0 = 0.01$ and
$\phi(z_1)$ is a randomly fluctuating phase with values in the
range $[0,2 \pi)$. Since there is not a radiation seed, we assume
$\bar\delta=0$.

\begin{figure}[h]
\begin{center}
\includegraphics[width=14cm]{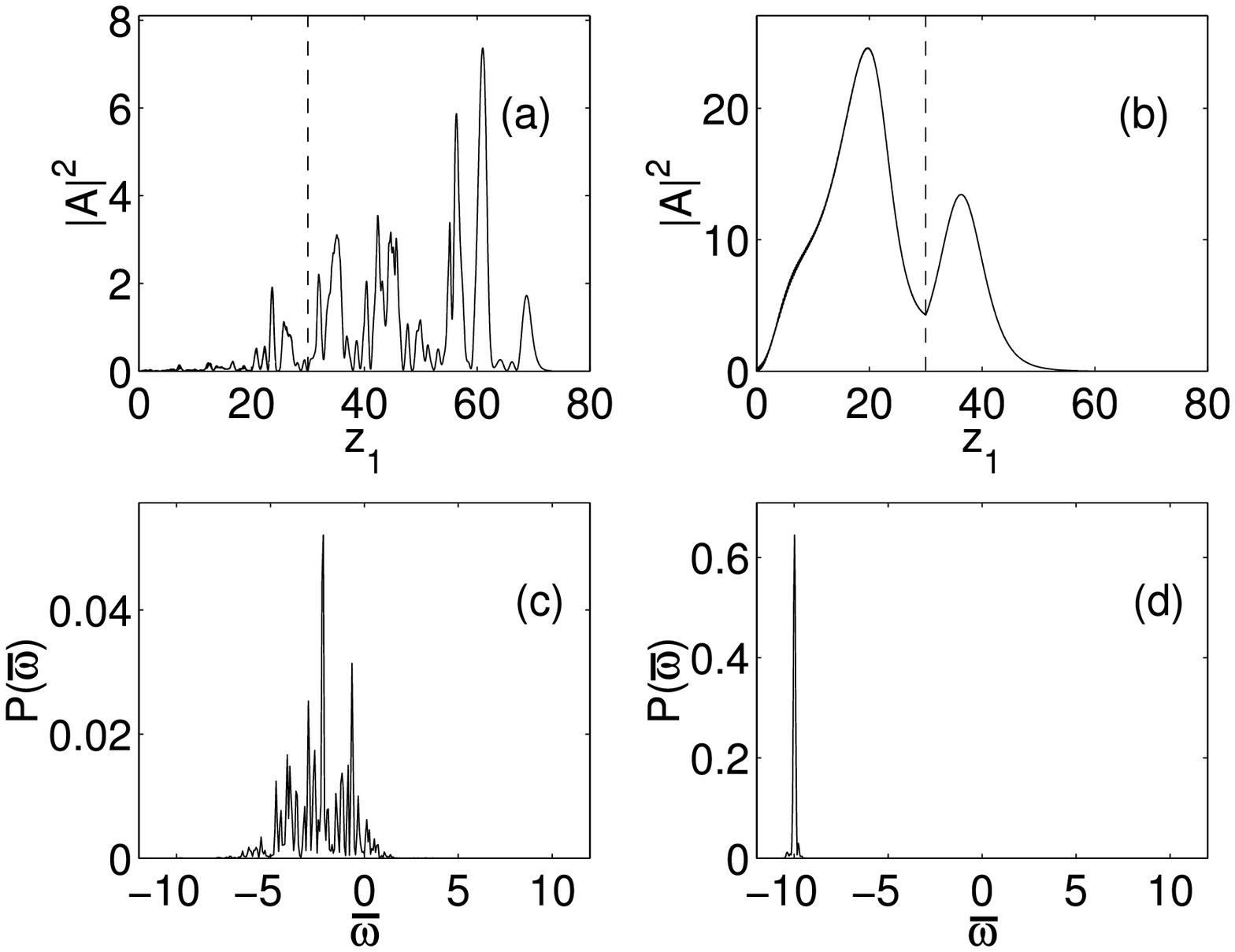}
\end{center}
\caption{Numerical solutions of eq.~({\ref{cn1})} and (\ref{cn2})
in the classical and quantum regimes : Graphs (a) and (b) show the
scaled intensity $|A|^2 (z_1)$ at $\bar{z}=50$ when the system
evolves classically ($\bar\rho=5$) and quantum mechanically
($\bar\rho=0.05$) respectively. Graphs (c) and (d) show the
corresponding scaled power spectra $P(\bar{\omega})$ at
$\bar{z}=50$ when $\bar\rho=5$ and $\bar\rho=0.05$ respectively.
The frequency shift in (d) is $1/2 \bar{\rho} = 10$ in agreement
with that predicted from fig.~\ref{fig1}. In all cases,
$\bar{L}=30$ and $\delta=0$ has been used. } \label{fig2}
\end{figure}

Fig.2 shows the numerical solution for $L_b=30 L_c$ and $\bar
z=50$ for the classical regime ($\bar\rho=5$, left column) and the
quantum regime ($\bar\rho=0.05$, right column). Figures \ref{fig2}a
and \ref{fig2}b show the intensity $|A|^2$ as a function of the
dimensionless variable $z_1$ (i.e. the coordinate along the
electron bunch in units of $L_c$ in the electron rest frame),
whereas figures \ref{fig2}c and \ref{fig2}d show the power
spectrum $P(\bar\omega)$ as a function of
$\bar\omega=(L_c/c)(\omega'-\omega)$, where $\omega'$ is the field
frequency and $\omega$ is the carrier resonant frequency. In fig.
\ref{fig2}a and \ref{fig2}b , $z_1=0$ is the trailing edge and
$z_1=30$ is the leading edge of the electron bunch. Therefore, the
region on the left of the dotted line is the radiation on the
electron beam, on the right is free propagation in vacuum.

\begin{figure}[h]
\begin{center}
\includegraphics[width=14cm,clip=true]{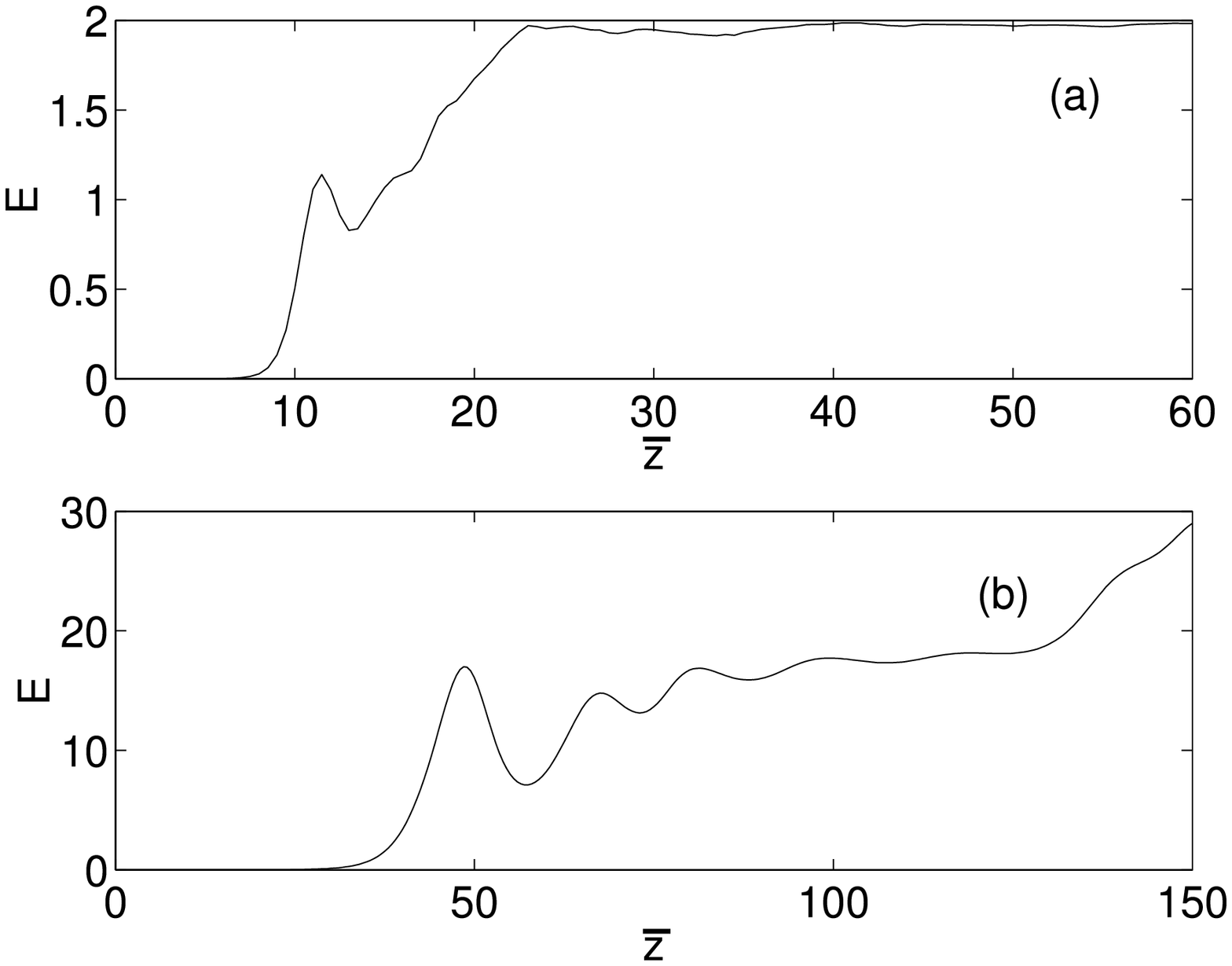}
\end{center}
\caption{Scaled energy, $E$, as a function of $\bar{z}$ when (a)
$\bar\rho=5$ and (b) $\bar\rho=0.05$. In both cases $\delta=0$ and
$\bar{L}=30$.} \label{fig3}
\end{figure}

Alternatively one can interpret fig. \ref{fig2}a and \ref{fig2}b
as the temporal behavior of the intensity for an observer which is
at a given position in the wiggler and which will see the pulse as
it appears from right to left in the figure. We stress that in the
classical theory with universal scaling the two cases should be
identical. On the contrary, the dramatic difference is evident
from  fig. \ref{fig2}: the temporal structure in the classical
limit (fig.~\ref{fig2}a) is almost chaotic and the width of the
spectrum is large (fig.~\ref{fig2}c). Conversely, the temporal
behavior in the quantum limit (fig.~\ref{fig2}b) shows a
purification of the initially noisy evolution so that the temporal
structure looks similar to the self similar solution one would
obtain with a coherent seed signal. The corresponding spectrum
becomes extremely narrow (fig.~\ref{fig2}d), much sharper than the
classical one. It can be seen in fig.~\ref{fig2}d that the
frequency is shifted by $1/2 \bar\rho=10$, in agreement with the
predictions of the linear analysis described earlier (see
fig.~\ref{fig1}). The behaviour of the system is similar to what
would be expected if the quantum cooperation length is much larger
than the classical cooperation length, so that all electrons
radiate coherently in the quantum regime. The difference between
classical and quantum SASE behavior is confirmed by a comparison
of fig.~\ref{fig3}a and \ref{fig3}b, which show the total radiated
energy, $E$ (in units of $|A|^2$), in the classical and quantum
regimes respectively, where
\[
E=\int_{0}^{\bar{z}+\bar{L}} |A|^2  d z_1.
\]
In the quantum regime we observe a behavior similar to that which
one would obtain with a coherent seed signal in the long bunch
superradiant case \cite{FEL:SR,NC}.

Finally, we discuss the reason for quantum purification of the
spectrum. As remarked earlier, in fig~\ref{fig1} the gain
bandwidth (which is the reciprocal of the ``real'' coherence
length) decreases as $\bar\rho^{3/2}$ in the quantum regime. Note
that the cooperation length in the quantum regime,
$L'_c=L_c/\sqrt{\bar\rho}$, has the same dependence on $\bar\rho$.
Hence, one can understand that in quantum SASE, i.e., for small
value of $\bar\rho$, the system behaves as if the startup of the
FEL interaction is initiated by a coherent bunching or coherent seed.

In conclusion, in this letter we have given a proof of principle
of the novel regime of Quantum SASE, with dynamical properties
very different from ``normal'' classical SASE. In particular,
quantum SASE predicts quantum purification of the temporal
structure and spectrum. The possibility of experimental
observation of this quantum regime is under investigation and will
be discussed elsewhere.

We acknowledge useful discussions with S. Bertolucci, L. Serafini,
M. Ferrario and L. Palumbo. This work is supported by INFN.

\end{document}